\documentstyle[12pt]{article}
\begin{document}
\begin{flushright} {OITS 697}\\
March 2001
\end{flushright}
\vspace*{1cm}
 
\begin{center} {\Large {\bf The effects of gluon depletion on
$J/\psi$ suppression \\in $pA$ and $AA$ collisions}}
\vskip .75cm
 {\bf  Rudolph C. Hwa$^1$, J\'{a}n Pi\v{s}\'{u}t$^2$ and Neva
Pi\v{s}\'{u}tov\'{a}$^2$ }
\vskip.5cm

{$^1$Institute of Theoretical Science and Department of Physics\\
University of Oregon, Eugene, OR 97403-5203, USA\\
\bigskip
$^2$Department of Physics, Comenius University, SK-84215,
Bratislava, Slovakia\\}
\end{center}

\begin{abstract}

The enhanced suppression of $J/\psi$ production at large $x_F$
in $pA$ collisions is studied in the framework of gluon
depletion at large $x_1$.  The nonperturbative process that
modifies the gluon distribution as the gluons propagate in
nuclear matter is described by an evolution equation with a
kernal to be determined by phenomenology. With nuclear
shadowing and anti-shadowing taken into account, the effect on
the gluon distribution is shown to be a depletion in excess of
40\% at $x_1 \approx 0.8$ for $A > 100$.  There is a small
amount of enhancement of the gluon distribution at small
$x_1$, but it does not lead to any contradiction with the
existing data on $J/\psi$ suppression in the central region. 
Extentions to $\psi^{\prime}$ suppression and $AB$ collisions
are also investigated in the framework of gluon redistribution.
\end{abstract}

\section{Introduction} 

In an earlier paper \cite{hpp} we
presented the phenomenological evidence for the depletion of
high-momentum gluons as a projectile proton traverses a target
nucleus in
$pA$ collisions.  Here we present a more complete discussion of
the subject with more technical details and with extension to
nucleus-nucleus collisions.

Our approach is unconventional in that we do not make the
usual assumption that the parton distributions in a proton
remain unaltered as the proton propagates through a nucleus,
even if a hard subprocess occurs deep in the nucleus.  That
assumption amounts to factorization, a property that has been
proven for $pp$ collision, but clearly must fail for $pA$
collision if $A$ is infinitely large.  For realistic nuclear
sizes our alternative assumption that the parton distribution
can be modified led to $J/\psi$ suppression
\cite{hpp2} that cannot be distinguished from the effects of the
usual mechanisms \cite {rv,ks}.  What is different now is the
appearance of new data on
$J/\psi$ suppression in $pA$ collisions at large $x_F$
\cite{ml}.  Since those data cannot be explained in terms of
hadronic absorption of the produced $c\bar{c}$ state
\cite{hhk,aga,rv2}, we find more direct support for the idea of
gluon depletion before the hard subprocess that produces the
$c\bar{c}$ state
\cite{hpp}. 

The data of the Fermilab E$866$ experiment \cite{ml}
on the
$J/\psi$ suppression in $pA$ collisions at $800$ GeVc are
expressed as the ratio
\begin{eqnarray}
R\left(x_F,A\right)=\sigma_A\left(x_F\right) /
A\sigma_N\left(x_F\right)=A^{\alpha\left(x_F\right)-1}
\quad,
\label{1}
\end{eqnarray}
where $\sigma_{N,A}$ is the cross section for
$J/\psi$ production by a proton on a nucleon $(N)$ or on a nucleus
$(A)$.  An analytical formula for $\alpha (x_F)$ is given in
\cite{ml}
\begin{eqnarray}
\alpha\left(x_F\right) = 0.96
\,\left(1-0.0519{x_F}-0.338{x^2_F}\right)\quad,
\label{2}
\end{eqnarray}
for $-0.1<{x_F}<0.9$. Equation (\ref{2}) differs
significantly from the one given in their preprint \cite{ml2},
on the basis of which the analysis in \cite{hpp} was done. 
Here we perform a reanalysis based on the new
parameterization in Eq.~(\ref{2}).

In our view there are three sources that can contribute to the
$x_F$ dependence, and we write them in a product form
\begin{eqnarray}
R\left(x_F, A\right) =  G\left(x_F, A\right)
N\left(x_F, A\right) H\left(x_F, A\right)
\quad,
\label{3}
\end{eqnarray}
where $H(x_F,A)$ represents the hadronic
absorption of the $c \bar{c}$ state, $N(x_F,A)$ nuclear
shadowing, and
$G(x_F,A)$ the gluon depletion effect, defined as
\begin{eqnarray}
G\left(x_F\left(x_1\right),A\right)=g\left(x_1,A\right) /
g\left(x_1,0\right)
\quad,
\label{4}
\end{eqnarray}
$x_1$ being the momentum fraction of the gluon in the projectile
proton, and $g(x_1,A)$ being the effective gluon distributions of
that projectile in a nucleus A at the point of $c\bar c$
production, averaged over the penetration depth at which the
hard process occurs.

By $H(x_F,A)$ we mean the
absorption factor that operates between the production of the
$c\bar{c}$ state and the detection of the final $J/\psi$ due to
any mechanism, including the interaction with comovers.  Since no
good arguments have been advanced to show that $H(x_F,A)$ can
have a significant dependence on $x_F$ \cite{hhk,aga,rv2}, we
shall assume in the following that $H(x_F,A)$ is independent
of
$x_F$.  This is not a serious limitation in our formalism.  If
a reliable
$x_F$ dependence is found at a later date, we can easily
incorporate it in our analysis.  For now we adopt the usual
Gerschal-H\"ufner form.
\begin{eqnarray}
H(A) = \exp \left[-\sigma \rho L (A) \right]
\label{5}
\end{eqnarray}
where $\sigma$ is the absorption cross section, $\rho$ the
nuclear density, and $L(A)$ the mean path length in $A$ that a
$c\bar{c}$ state propagates.

In the next section the nuclear shadowing factor $N(x_F,A)$
will be discussed.  We shall find a simple formula that can
represent the change in the gluon distribution in the target
nucleus due to shadowing and anti-shadowing.  Such a formula
facilitates the analysis, and offers a simple parametrization
that is convenient to use, independent of the particular
problem that we apply it to here.

The determination of the depletion factor $G(x_F (x_1),A)$ is
the main theme of this paper.  In Sec.~3 we shall go
beyond a review of the content of Ref.~\cite{hpp} , not
only because the data has changed from those in the
preprint \cite{ml2}, resulting in numerical differences,
but also because we shall improve on \cite{hpp} in some
technical details and include some new material.

In Sec.~$4$ we shall extend the result from the study of
the $pA$ problem to $AB$ collisions.  We shall show how the
enhanced depletion at large $x_F$ does not affect the $J/\psi$
suppression at small $x_F$, which is where the existing data
for $AB$ collisions were collected.  We shall have nothing to
add on the subject of enhanced suppression observed in
$Pb$-$Pb$ collisions beyond what we have advanced in
Ref.~\cite{hpp2}.

\section {Nuclear Shadowing and Anti-shadowing}

The nuclear shadowing and anti-shadowing problem has been
studied phenomenologically by Eskola {\it et.~al.}
\cite{ekr,eks}.  Instead of focusing on the physics of the
origin of the problem in QCD, they analyzed the deep
inelastic scattering data of nuclear targets.  On the basis
of DGLAP evolution
\cite{yd} they can determine the parton distribution at
any $Q^2>2.25$ GeV$^2$.  The results are given in terms of
numerical parametrizations (called EKS98 \cite{eks}) of
the ratio
\begin{eqnarray}
N_i^A (x,Q^2) = f_{i/A}(x,Q^2)/ f_i (x,Q^2),
\label{6}
\end{eqnarray}
where $f_i$ is the parton distribution of flavor $i$ for
the free proton and $f_{i/A}$ is that for a proton in a
nucleus $A$.

For the purpose of $J/\psi$ production we are interested in
Eq.\ (\ref{6}) for only the gluons.  We consider only the
dominant subprocess
$g_1(x_1) + g_2(x_2) \rightarrow c+\bar{c}$, where $x_1$ is
the momentum fraction of the projectile gluon whose
distribution is $g_1(x_1)$, and $x_2$ is that of the target
gluon whose distribution is $g_2(x_2)$.  For the $c\bar{c}$
state produced with momentum fraction $x_F=x_1-x_2$, the
usual kinematical relations are
\begin{eqnarray}
x_{1,2}=  [(4 \tau + x_F)^{1/2} \pm x_F] /2,
\hspace{1cm} x_1x_2=\tau \equiv M^2_{J/\psi}/s, 
\label{7}
\end{eqnarray}
where it is assumed that the $c\bar{c}$ state that turns to
$J/\psi$ is produced near threshold.  Thus the virtuality of
the subprocess $g+g \rightarrow c + \bar{c}$ is given by the
value of $M^2_{J/\psi}$, or slightly higher.  We shall take
$Q^2=10$ GeV$^2$, a value that is chosen in EKS98 \cite{eks} to
give explicit values of ${N_i}^A(x_1,Q^2)$.  For simplicity we
shall label the gluon distribution at $Q^2=10$ GeV$^2$ by
$N(x_2,A)$.

The numerical output of EKS98 for $N(x_2,A)$ is shown by
the points in Fig.~1 for $A=50$, $100$, and $200$.  We
exhibit only the values for $x_2$ in the range
$0.01 \leq x_2 \leq 0.12$, since that is the range relevant for
the production of $J/\psi$ at $800$ GeV/c for $0<x_F<0.8$.  For
the purpose of convenience in using those values of
$N(x_2,A)$ in analytic manipulation and computation, we
propose a simple formula that contains the shadowing and
anti-shadowing effects.  Since the cross-over of the two
effects occurs at $x_2 = 0.02$ where $N(x_2,A) = 1$ for all $A$,
it is sensible to use an auxiliary variable $\xi$, defined by
\begin{eqnarray}
\xi=3.912 + {\rm ln} x_2
\label{8}
\end{eqnarray}
which vanishes at $x_2=0.02$.  Moreover, we notice that ${\rm
ln} N(x_2,A)$ depends linearly on ${\rm ln} A$ to a good
approximation, so it suggests a power-law behavior
\begin{eqnarray}
N(x_2,A)=A^{\beta(x_2)}.
\label{9}
\end{eqnarray}
The exponent $\beta(x_2)$ can be determined by fitting the
data for $A=100$.  In terms of $\xi$ we find a good fit with
the parametrization
\begin{eqnarray}
\beta(\xi)= \xi(0.0284 + 0.0008 \xi-0.0041 \xi^2),
\label{10}
\end{eqnarray}
the result of which is shown in Fig.~2, where the points
are for ${\rm ln} N(x_2,A)/\ln A$ obtained from EKS98 at
$A=100$.  This convenient formula for $\beta(\xi)$ can then
be used in conjunction with Eq.\ (\ref{8}) to determine
$N(x_2,A)$ for other values of $A$.  The curves in Fig.~1
exhibit the good agreement between the data and our
parameterization for $A=50$, $100$, and $200$.

In the following we shall simply use Eq.\ (\ref{9}) as a
summary of the effects of nuclear shadowing and
anti-shadowing for problems in $pA$ collisions where the
relevant range of $x_2$ is in the interval $0.01 \leq x_2 \leq
0.12$.

\section{Evolution of Gluon Distribution in a Nucleus}

We now may regard $R(x_F,A)$ and $N(x_2,A)$ as known
phenomenologically.  Thus, from Eq.\ (\ref{3}) we may write 
\begin{eqnarray}
G(x_F,A) H(A)=A^{\alpha(x_F)-\beta(x_2(x_F))-1}.
\label{11}
\end{eqnarray}
Although the form of $H(A)$ is given by Eq.\ (\ref{5}), we do
not know the value of $\sigma$ in the present circumstance
where we allow the possibility of gluon depletion.  Hence, we
treat $H(A)$ temporarily as unknown, along with the depletion
factor $G(x_F,A)$.  However, the $x_F$ dependence is completely
known from the RHS of Eq.\ (\ref{11}).

To proceed we need some theoretical input on the possible form
of $G(x_F,A)$, or more directly the gluon distribution
$g(x_1,z)$ in the projectile, where $z$ is the distance
traversed in a nucleus.  Note that we have refrained from
referring to the projectile as the proton, since the possible
modification of $g(x_1,0)$ for $z>0$ implies that the incident
proton loses its usual identity, in particular, the nature of
its partonic content, as the projectile, now identified only as
a flux of partons, propagates in the nuclear medium.  How
$g(x_1,z)$ evolves in the nuclear medium is clearly a
nonperturbative process that involves multiple scatterings of
gluons and quarks at low virtualities.  Nevertheless, for every
incremental distance, $dz$, that a gluon travels the
modification that $g(x_1,z)$ undergoes must be perturbative in
that $g(x_1,z+dz)-g(x_1,z)$ is small and is proportional to
$dz$.  It is therefore reasonable to adopt an evolution
equation similar in spirit to that of DGLAP \cite{yd}, but
with the change in resolution scale $d {\rm ln} Q^2$ replaced
by the change in penetration depth $dz$, so that we write
\begin{eqnarray}
\frac{d}{dz} g(x,z)=\int_{0}^{1}\frac{dx^{\prime}}{x^{\prime}}
g(x^{\prime},z)Q(\frac{x}{x^{\prime}}),
\label{12}
\end{eqnarray}
where the unknown kernel $Q(x/x^{\prime})$ controls the gain
and loss of the gluons in $dz$.  $Q(y)$ cannot be determined by
perturbative calculation, as the splitting functions in pQCD
for $Q^2$ evolution.  Equation (\ref{12}) is similar to the
nucleonic evolution equation proposed in \cite{rh}, except
that this is now at the parton level.  In Eq.\ (\ref{12}) the
quark sector has been left out for simplicity.  To be more
complete one should include also the effects of the couplings of
gluons with the quarks, a task that is deferred to the future. 
Thus what we can achieve now is the determination of
an effective kernel $Q(y)$ that can account for the enhanced
suppression of
$J/\psi$ at large $x_F$.

To determine the $z$ dependence of $g(x,z)$ let us take the
moments by defining
\begin{eqnarray}
g_n(z)=\int_{0}^{1}dx \, x^{n-2}g(x,z)
\label{13}
\end{eqnarray}
and
\begin{eqnarray}
Q_n=\int_{0}^{1}dx \, x^{n-2} Q(y).
\label{14}
\end{eqnarray}
Then by the convolution theorem, Eq.\ (\ref{12}) becomes 
\begin{eqnarray}
dg_n(z)/dz=g_n(z)Q_n,
\label{15}
\end{eqnarray}
whose solution is
\begin{eqnarray}
g_n(z)=g_n(0)e^{zQ_n}.
\label{16}
\end{eqnarray}
It is possible that Eq.\ (\ref{15}) is valid only when $z$ is
large enough, in which case Eq.\ (\ref{16}) should be
modified to read
\begin{eqnarray}
g_n(z)=g_n(z_0)e^{(z-z_0)Q_n}
\label{17}
\end{eqnarray}
for $z$ greater than some positive value of $z_0$.  

To proceed,
let us substitute Eq.\ (\ref{4}) in (\ref{11}) and define 
\begin{eqnarray}
J(x_1,A)\equiv g(x_1,0)A^{\alpha(x_F(x_1))-\beta(x_2(x_1))-1}
\label{18}
\end{eqnarray}
where the interrelationships among $x_1$,$x_2$ and $x_F$ are
specified by Eq.\ (\ref{7}).  For $g(x_1,0)$ we use the
canonical form
\begin{eqnarray}
g(x_1,0)=g_0(1-x_1)^5.
\label{19}
\end{eqnarray}
The final result is insensitive to its form and independent of
$g_0$, which we shall set to be $1$.  Thus we may regard
$J(x_1,A)$ as known.  Since we also have
\begin{eqnarray}
J(x_1,A)=g(x_1,A)H(A),
\label{20}
\end{eqnarray}
its moments are, by virtue of Eqs.\ (\ref{5}) and (\ref{16}),
\begin{eqnarray}
J_n(z)=g_n(0) {\rm exp} [z(Q_n-\sigma\rho)].
\label{21}
\end{eqnarray}
Here and in the following we shall use $z$ (until Sec. 5) to
denote the average penetration depth (i.e., $z\equiv \bar z_A$)
in $A$ when a
$c\bar{c}$ state is produced by
$gg$ annihilation.  It is then also the average length that the
$c\bar{c}$ must travel in $A$ and be subject to hadronic
absorption, i.e., $z=L(A)$.

We can determine $J_n(z)$ by taking the moments of $J(x_1,A)$,
as expressed in Eq.\ (\ref{18}).  However, there is a problem
in evaluating $J_n(z)=\int_{0}^{1}dx_1{x_1}^{n-2}J(x_1,z)$,
since $J(x_1,A)$ is ill-defined at $x_1=0$.  According to
Eq.\ (\ref{7}), $x_2$ diverges as $x_1\rightarrow 0$; thus
limiting $x_2$ to $1$ implies that $x_1$ cannot be less than
$M^2_{J/\psi}/s$, a small but nonvanishing value. 
Furthermore, $x_F$ becomes negative at small $x_1$, and we
lose any knowledge about $\alpha(x_F)$ for $x_F <- 0.1$
\cite{ml}.  Also, $\beta (x_2)$ is not reliably known at large
$x_2$, so the RHS of Eq.\ (\ref{18}) cannot offer accurate
determination of $J(x_1,A)$ as $x_1\rightarrow 0$.  These
defects can be suppressed by the factor ${x_1}^{n-2}$ in the
integrand, if we restrict $n$ to
$\geq 3$.  We shall therefore determine $J_n$ only for $n \geq
3$.

For convenience, let us define
\begin{eqnarray}
K_n(z)\equiv {\rm ln} [J_n(z)/g_n(0)]=z(Q_n-\sigma\rho).
\label{22}
\end{eqnarray}
Using Eqs.\ (\ref{2}), (\ref{10}) and
(\ref{19}) in
(\ref{18}), we can calculate $J_n(x)$ and therefore $K_n(z)$
for $n
\geq 3$.  As mentioned earlier, Eq.\ (\ref{2}) is different
from a previous form of $\alpha (x_F)$ given in \cite{ml2} and
used in
\cite{hpp}.  The results are shown as discrete points in
Fig.~3 for
$A=100$ and $200$.  The corresponding values of $z$ are halves of
the average total path lengths of the nuclei, i.e.,
$z = 3R_A/4 = 0.9A^{1/3}$ fm  and are therefore $z_1
= 4.177$  and $z_2 = 5.262$ fm, respectively.

To extract the information contained in those points in
Fig.~3 we need an analytical representation of $K_n$.  We
choose to fit $K_n$ by the following formula, different from
the one used in \cite{hpp},
\begin{eqnarray}
K_n=\sum_{i=0}^{3} k_i n^i + k_4 n^{1/2}.
\label{23}
\end{eqnarray}
The results of our fits are shown by the smooth curves in
Fig.~3.  The corresponding parameters are given in Table
1.  Note that the fits allow us to extrapolate smoothly to
$n=2$, where we could not calculate $J_2$.

\begin{table}
\begin{center}
\caption{Values of the coefficients $k_i$}
\vspace{.5cm}
\begin{tabular}{|c|c|c|c|c|c|c|}
\hline
 A & z & $k_0$ & $k_1$ & $k_2$ & $k_3$ & $k_4$ \\ \hline
 $100$ & $4.177$ & $-0.5444$ & $-0.2063$ & $5.87 \times 10^{-3}$ &
$-8.35 \times 10^{-5}$ & $0.424$ \\ \hline
$200$ & $5.262$ & $-0.6267$ & $-0.2354$ & $6.64 \times 10^{-3}$ &
$-9.39 \times 10^{-5}$ & $0.486$ \\
\hline
\end{tabular}
\end{center}
\end{table}

Before we determine $Q_n$ from $K_n$ in Eq.\ (\ref{22}), we
need to verify the $z$ dependence.  It should first be
recognized that the experimental parameterization of the $A$
dependence, such as in Eq.\ (\ref{1}), is not compatible with
the theoretical expectation, such as in Eq.\ (\ref{5}) and
(\ref{21}), except in a certain range of $A$.  Since a
power-law
$A^\gamma$, expressed as $e^{\gamma {\rm ln} A}$, is
approximately
$e^{\gamma^{\prime} z}$, where $z \propto A^{1/3}$, only when
${\rm ln} A$ is approximately $A^{1/3}$, the correspondence
can only be for
$60<A<240$.  With that understanding, let us nevertheless
calculate $J(x_1,z)$ for all $z<6$ using $A=(z/0.9)^3$ in
Eq.\ (\ref{18}), take the moments, and then determine $K_n(z)$
through the first half of Eq.\ (\ref{22}).  The result is
shown as points in Fig.~4 for eleven values of $z$ between
$0.9$ and $5.9$, corresponding to $A$ from 1 to 282, and for
four representative values of $n$, viz., 3, 8, 13 and 20.  The
straight lines are linear fits of the last six points for each
value of
$n$.  Evidently, the $z$ dependence of $K_n(z)$ is very nearly
linear for $3.4<z<5.6$, which corresponds to $54<A<240$.  Thus
our theoretical formalism is consistent with the
experimental data in the region where $lnA \approx A^{1/3}$. 
At $z=0.9$, or $A=1$, all points converge to $K_n=0$, as they
should.  We cannot reliably apply our formalism to the
collision problems where $A<50$.  Fig.~4 also suggests that
even when $A$ is large, say $>100$, the gluon evolution
equation (\ref{12}) may not be valid at small $z$, here used
in the sense of penetration depth within the large nucleus,
not the average depth.  In the following we shall limit our
consideration to only the linear portion of Fig.~4.  In that
region the second half of Eq.\ (\ref{22}) should be treated
as differentially correct, i.e.,
\begin{eqnarray}
\Delta K_n (z)/ \Delta z =Q_n-\rho \sigma
\label{24}
\end{eqnarray}
where $\Delta K_n (z)= K_n(z+\Delta z)-K_n(z).$

Since the values of $k_i$ in Table 1 are determined in the
linear region, we can use them to obtain $Q_n$.  If we write
\begin{eqnarray}
Q_n= \sum_{i=0}^3 q_i n^i +q_4 n^{1/2},
\label{25}
\end{eqnarray}
then we have from Eqs.\ (\ref{23}) and (\ref{24})
\begin{eqnarray}
q_0=\Delta k_0/\Delta z + \rho \sigma , \quad \quad \
q_i=\Delta k_i/\Delta z \ (1\leq i \leq 4).
\label{26}
\end{eqnarray}
For the two $z$ values in Table 1, we get (with $\Delta z =
1.086$ fm)
\begin{eqnarray}
\begin{array}{lll}
q_0=-0.0758 + \rho \sigma, & q_1=-0.0268,   \\ 
q_2=7.13 \times 10^{-4}, & q_3=-9.6 \times 10^{-6}, & q_4=0.0568
\label{27}
\end{array}
\end{eqnarray}
in units of fm$^{-1}$.

Since the absorption cross section $\sigma$ is unknown when
gluon depletion is not negligible, $q_0$ is not fixed by
Eq.\ (\ref{27}).  Whatever the dynamics of gluon depletion
is, we require that the total gluon momentum does not increase
with
$z$.  Since the gluon momentum is $\int dx g(x,z)=g_2(z)$, that
requirement implies in conjunction with Eq.\ (\ref{17}) that
$Q_2 \leq 0 $.  Choosing the upper bound $Q_2=0$ leads to the
condition, on account of Eqs.\ (\ref{25}) and (\ref{27}),
\begin{eqnarray}
q_0=-(2q_1+4q_2+8q_3+\sqrt{2} q_4)=-0.0295.
\label{28}
\end{eqnarray}
With all parameters in Eq.\ (\ref{25}) now determined, the
$n$ dependence of $Q_n$ can be exhibited, as shown in
Fig.~5.  Evidently, $Q_n$ is smoothly varying and the use of
the polynomials in Eqs.\ (\ref{23}) and (\ref{25}), which are
different from those used in \cite{hpp}, is justified.

Although it is more direct to proceed immediately to the use of
Eq.\ (\ref{17}) to the determination of $g_n(z)$ and
therefore $G(x_1,z)$, which is our goal, there is some
advantage in an attempt to find $Q(y)$ at this point, while
we are on the subject of $Q_n$.  The easiest way to do that is
to put $Q_n$ in form
\begin{eqnarray}
Q_n=c_0+ \sum_{j} c_j/(n+j-1) \ ,
\label{29}
\end{eqnarray}
used in \cite{hpp}, so that it can imply directly
\begin{eqnarray}
Q(y)=c_0 \delta (1-y) + \sum_{j} c_j y^i \ ,
\label{30}
\end{eqnarray}
To translate from Eq.\ (\ref{25}) and (\ref{29}), we fit
the values of $Q_n$ at the 19 integer points, $2 \leq n
\leq 20$, determined by (\ref{25}), by use of the formula
(\ref{29}) with a suitable number of terms in the sum.  It
turns out that a good fit can be achieved with three terms:
$j = 3, 4$ and $5$.  The result is
\begin{eqnarray}
c_0=-0.1988, \quad c_3=6.205, \quad c_4=-23.316,\quad 
c_5=19.866
\quad .
\label{31}
\end{eqnarray}  
We show in Fig.~6 both the discrete values of $Q_n$ at
integral $n$ and the fitted curve using Eqs.\ (\ref{29}) and
(\ref{31}).  The corresponding function $Q(y)$, calculated
using Eqs.\ (\ref{30}) and (\ref{31}), is shown in Fig.~7.  It
is evident that the $c_0$ and $c_4$ terms correspond to gluon
depletion, while the $c_3$ and $c_5$ terms correspond to
gluon regeneration.

To determine $G(x_1,z)$ as defined in Eq.~(\ref{4}), we have  
Eq.~(\ref{17}) that specifies the evolution in $z$ in the
linear region (see Fig.~4) from $z_0$.  At this point we have
no formalism to extrapolate in the nonlinear region from
$z_0$ down to $0$, and it is in reference to $g(x_1, 0)$
that $G(x_1,z)$ is defined.  However, in view of the fact
that the hadronic absorption term $H(A)$ is known empirically
to be an exponential, as in Eq.~(\ref{5}), i.e., ${\rm ln}
H(z)$ being linear in $z$ for all $z$, one may regard the
nonlinear portion of Fig.~4 to be due primarily to the
mismatch between
${\rm ln} A$ and $A^{1/3}$ at low $A$.  Then we adopt the
approximation that Eq.~(\ref{16}) is adequate for relating
$g_n(z)$ to $g_n(0)$.  With $Q_n$ being known from
Eqs.~(\ref{25}), (\ref{27}) and (\ref{28}), and $g_n(0) = B
(n - 1, 6)$ which is the beta function, we can calculate
$g_n(z)$.  The result is shown in Fig.~8 by the full (open)
circles for $A = 100\,(200)$,  It is natural to fit the
resultant $g_n(z)$ by a linear combination of beta functions
in the form
\begin{eqnarray}
g_n(z) = \sum^3_{i = 1} a_i (z) B (n - 1, 5 + i )
\quad ,
\label{32}
\end{eqnarray}
The coefficients $a_i(z)$ are determined by fitting
$g_n(z)/g_n(0)$ in order to reduce the range of variation. 
For the two values $z_1$ and $z_2$, corresponding to $A =
100$ and $200$, we obtain
\begin{eqnarray}
\begin{array}{lll}
a_1(z_1 ) = 0.3526,&a_2(z_1) =
1.44,&a_3(z_1) = - 0.78,\\
a_1(z_2 ) = 0.2362,&a_2(z_2 ) =
1.655,&a_3(z_2 ) = - 0.869.
\end{array}
\label{33}
\end{eqnarray}
The curves in Fig.~8 are generated using Eqs.~(\ref{32}) and
(\ref{33}).  Evidently, the fits are good.

The inverse transform of the moments in Eq.~(\ref{32}) is
\begin{eqnarray}
g (x_1, z ) = \sum^3_{i = 1} a_i (z) (1 - x_1)^{4+i} \quad ,
\label{34}
\end{eqnarray}
whose implication for 
\begin{eqnarray}
G(x_1, z) &=& g (x_1, z ) /g (x_1, 0)\nonumber \\ 
&=&a_1(z) + a_2 (z) (1 - x_1) + a_3 (z) (1-x_1)^2
\label{35}
\end{eqnarray}
can readily be calculated using Eq.~(\ref{33}).  The results for
$z_1$ and $z_2$ are shown in Fig.~9.  Clearly, there is
significant depletion of gluons at large $x_1$, roughly 40\%
at $x_1 \sim 0.8$.  There is a small amount of regeneration at
small
$x_1$.  The cross-over is at around
$x_1 \approx 0.2$.  Although the enhancement at small $x_1$,
is at the 2 to 3\% level, in terms of the number of gluons in
a small $dx$ interval it is not insignificant compared to the
depletion at large $x_1$, because $g(x_1, 0)$  is strongly
damped at large $x_1$.  That is how the condition
$Q_2 = 0$ is satisfied.  

\section{Suppression of $J/\psi$ and $\psi^{\prime}$}

Having determined $G(x_1, A)$ in the previous section, we can
now return to the problem of charmonium suppression,
including $\psi^{\prime}$.  The Fermilab E866 experiment
\cite{ml} gives data for both $J/\psi$ and $\psi^{\prime}$ in
the form of $\alpha (x_F)$.  In using Eq.~(\ref{3}) to
calculate $R(x_F, A)$, and then $\alpha (x_F)$, we have 
Eqs.~(\ref{35}), (\ref{9}) and (\ref{5}) for $G(x_1, A)$,
$N(x_2, A)$ and $H(A)$, respectively.  For both $J/\psi$ and
$\psi^{\prime}$ we use $\sigma = 6.5$ mb in $H(A)$.  The
results on $\alpha (x_F)$ are shown as curves in Fig. 10 for
$J/\psi$ and Fig. 11 for $\psi^{\prime}$.  The two curves in
each figure are for $A = 100$ and $200$; they are
sufficiently close to each other to be
almost independent of $A$, thereby affirming our gluon
evolution model for that range of $A$.  

The most significant part of what we have
learned from this work on the $pA$ data is that it is hard to
reproduce the strong damping of the measured $\alpha(x_F)$ at
large $x_F$ without a substantial amount of gluon depletion
at large $x_1$, as seen in Fig.\ 9.  Furthermore, if there is
significant modification of the parton distribution as the
penetration depth increases, then it is hard to justify the
notion that a proton can traverse a large part of the target
nucleus without any changes, or that it can be wounded upon
the first collision with a nucleon and then remaining
unchanged thereafter.

So far out attention has been given to the $x_F > 0$ region
only where data exist.  However, the $x_F < 0$ region has
interesting physics also, and the data can be obtained either
at RHIC, or at a fixed target experiment with proton being
the target for a heavy-ion beam.  At large negative $x_F,
\, x_1$ would be small where Fig. 9 shows a very small
enhancement, not depletion.  On the other hand, $x_2$ would be
large, where
$N(x_2, A)$ would exhibit not only anti-shadowing, but also the
EMC and Fermi motion effects.  As an illustration of the $x_F$
dependence, we have performed the calculation for
$J/\psi$ suppression in $pW$ collision at three energies and
for all $x_F$, negative as well as positive.  The absorption
cross section is set at $\sigma = 8$ mb, a value found in the
next section.  The values of $N(x_2, A)$ for the full $x_2$
range are obtained from EKS98 \cite{eks}.  The results are shown
in Fig. 12.  We see that the negative $x_F$ region shows very
little dependence on energy and reflects mainly the property of
$N(x_2, A)$.

Gluon depletion of the type discussed in this paper has no
effect on the suppression of $J/\psi$ in the 
$x_F < 0$ region.  However, there is another type of gluon
depletion, discussed in the second paper in Ref. \cite{hpp2},
called nonlinear depletion, that can influence the survival
probability in the $x_F < 0$ region.  When $x_2$ is large, the
fast gluons in the rear part of the nucleus can catch up and
interact with the slow gluons released by the $p$-$N$
collisions in the front part of the nucleus, leading to a
depletion of gluon at large $x_2$.  Although this involves the
interaction between gluons in different nucleons in the
nucleus, it is not nuclear shadowing, since the conventional
nuclear shadowing at low $x_2$ does not require the invasion
of an external proton to initiate the cascading interactions
of gluons among the broken nucleons.  We have no prediction on
the nature of the effect of this type of depletion.  If the
experimental data on the production rate differ significantly
from the curves in Fig. 12 in the $x_F < 0$ region, then there
will be strong motivation to consider this unusual type of
gluon depletion.  

\section{Nucleus-nucleus Collisions}

Having investigated the $pA$ collision problem above, and
finding the necessity to consider the depletion of gluons in
the projectile before the production of $c\bar{c}$ states in
the $x_F > 0$ region, it is natural to ask next what the
implication would be for the $J/\psi$ suppression problem in
$AB$ collisions.  Even without any detailed calculations it is
straightforward to infer that there will be enhanced
suppression at large $x_F$.  However, the currently available
data on $J/\psi$ suppression in nuclear collisions are from
CERN-SPS, and are limited to the central rapidity region.  It
is therefore our burden to show that the gluon depletion
mechanism discussed in Sec. 3 is consistent with the existing
nuclear collision data at small
$x_F$.

In Sec.~3 we have used
$z$ mainly as the average penetration depth in the target
nucleus where
$c\bar{c}$ is produced, although in the evolution equation
(\ref{12}) $z$ is the actual path length.   Since in this
section we need to average the production point over all impact
parameters in the $AB$ collisions, we now restore $z$ to be
actual path length of a gluon in a nucleus.  Thus we allow it
to vary from 1 to 12 fm.  We use Eq. (\ref{16}) to determine
$g_n(z)$ at all odd integer points in the range $1 \leq z \leq
12$ (for the sake of simplicity in fitting), and then fit them
at each such value of
$z$ by the formula (\ref{32}).  The resulting values of
$a_i(z)$ are shown by the points in Fig.~13, they are in turn
fitted by quadratic equations of the form
\begin{eqnarray} 
a_i(z) = b_{i0} + b_{i1}z + b_{i2}z^2 \quad .
\label{36}
\end{eqnarray} 
The result is shown by the curves in Fig.\ 13.  The
corresponding coefficients are
\begin{eqnarray}
\begin{array}{lll} 
b_{10} = 0.974, & b_{11} = -0.175, &b_{12} =
0.0068, \\ b_{20} = 0.0866, & b_{21} = 0.402, &b_{22} = -0.0202,
\\ b_{30} = -0.066, & b_{31} = -0.226, &b_{32} = 0.0142.
\end{array}
\label{37}
\end{eqnarray} 
Using Eqs. (\ref{36}) and (\ref{37}) in (\ref{34}), we
can evaluate
$g\left(x_1, z\right)$ at all $x_1$ and $z$.

The cross section for the production of $J/\psi$ in
$AB$ collisions can be calculated in the standard way.  We shall
just write it down as follows (see, e.g., Refs \cite{hpp2,rv}):
\begin{eqnarray}
\sigma_{J/\psi} &=& \int d^2b\, d^2s\, dz_A\, dz_B \, \rho_A
(\vec{s}, z_A) \rho_B (\vec{s}-\vec{b}, z_B)\nonumber\\
&&\cdot \int {dx_1 \over x_1} {dx_2 \over x_2}\, g_A (x_1, L_B-z_B)\,g_B (x_2,
L_A-z_A)\, N(x_1,A) N (x_2, B)\nonumber\\
&&\cdot e ^{- \sigma_a \left[\rho_A (L_A + z_A) + \rho_B (L_B
+ z_B)\right]} 
\hat{\sigma} _{gg \rightarrow c\bar{c}} (x_1, x_2)
\quad ,
\label{38}
\end{eqnarray}
where we have included the $z_{A,B}$ dependences in the gluon
distributions $g_{A,B} (x_{1,2}, L_{B, A}-z_{B, A})$
 and the nuclear shadowing functions $N(x_1, A)$ and $N(x_2,
B)$.  $L_{A,B}$ are the path length through $A(B)$
 at the distances $\vec{s}$ and $\vec{s}-\vec{b}$,
respectively, from the centers of the nuclei, i.e.
\begin{eqnarray}
L_A = (R_A^2 - s^2)^{1/2}\quad, \quad \quad L_B =  (R^2_B
-\left|\vec{s}- \vec{b} \right|^2)^{1/2}\quad .
\label{39}
\end{eqnarray}
The gluon distribution $g_{A} (x_{1}, L_{B}-z_{B})$ is given by
[see Eq. (\ref{34})]
\begin{eqnarray}
g_{A} (x_{1}, L_{B}-z_{B}) = \sum^3_{i = 1} a_i (L_B-z_B)
(1 - x_1)^{4+i} \quad ,
\label{40}
\end{eqnarray}
and similarly, for $g_B(x_2, z_A)$, the coefficients $a_i(z)$
in Eq.~(\ref{34}) are replaced by $L_A - z_A$, which is the
distance that a parton in $B$ travels in $A$ before the
production of $c\bar{c}$ at $z_A$.  The distance that a
$c\bar{c}$ state travels in $A$ is  
$L_A + z_A$, and $z_A$ is integrated from $-L_A$ to $+L_A$. 
For the energies at CERN-SPS, $\sqrt{s} \approx 18-20$ GeV,
and for $x_F \approx 0$, the hard cross section
$\hat{\sigma}_{gg \rightarrow c\bar{c}} (x_1, x_2)$ restricts
the gluon momentum fractions to $x_1 \approx x_2 \approx
M_{J/\psi}/ \sqrt{s} \approx 0.16$.

The survival probability is
\begin{eqnarray}
S^{AB}_{J/\psi} &=&
\sigma^{AB}_{J/\psi}/\sigma^{AB(0)}_{J/\psi}\nonumber \\ 
&=& N^{-1}_{AB} \int d^2b d^2s \int^{L_A}_{-L_A}dz_A
\int^{L_B}_{-L_B}dz_B W (\vec{b}, \vec{s}, z_A, z_B )
\label{41}
\end{eqnarray}
where
\begin{eqnarray}
W (\vec{b}, \vec{s}, z_A, z_B) &=& G_A (x_1, L_B-z_B)\,G_B
(x_2, L_A-z_A)\, N(x_1,A) N (x_2, B) \nonumber\\
&&\cdot e ^{- \sigma_a \left[\rho_A (L_A + z_A) + \rho_B (L_B
+ z_B)\right]} 
\quad ,
\label{42}
\end{eqnarray}
and $N_{AB}$ is the same integral in Eq.~(\ref{41}) but with
$W (\vec{b}, \vec{s}, z_A, z_B)$ replaced by 1.  $G_A (x_1,
L_B-z_B)$ is as given in Eq.~(\ref{35}) except that $z$ is
replaced by $L_B-z_B$.  Because of both the gluon enhancement at
$x_{1,2}\simeq 0.16$ and the anti-shadowing, the absorption
cross section
$\sigma_a$ now has to be somewhat larger than before
\cite{hpp2}.  An overall agreement with all the $pA$ and $AB$
collision data, except the $Pb$-$Pb$ case, can be achieved with
the use of one value of $\sigma_a = 8$ mb.  The result is
given in Table 2 for the various $AB$ cases.  Fig.~14 shows
 \begin{table}[h]
\begin{center}
\caption{Survival probability for various $AB$ collisions}
\vspace{.5cm}
\begin{tabular}{|c|c|c|c|c|c|c|c|c|c|}\hline
$AB$&$pp$&$pC$&$pAl$&$pW$&$pU$&$OCu$&$OU$&$SU$&$PbPb$\\ \hline
$S$&1&0.85&0.82&0.70&0.68&0.67&0.58&0.56&0.49\\ \hline
\end{tabular}
\end{center}
\end{table}
how those values compare with the experimental data \cite{mg}
by the straightline segments that connect those successive
points.  It is evident that apart from the $Pb$-$Pb$ case the
agreement with the data is satisfactory.  Thus we can
conclude that the gluon depletion mechanism used to treat
the $pA$ problem leads to no disagreement with the $AB$
collisions---except that the $Pb$-$Pb$ case remains as an
anomaly.

Finally, we compute the $x_F$ dependence of the $J/\psi$
suppression factor for just one nuclear-collision case as an
example, which we take to be $Pb$-$Pb$.  We consider two
energies:  $E_{\rm lab} = 160$ GeV and $\sqrt{s} = 60$ GeV
for RHIC.  Except for the kinematics in $\hat{\sigma}_{gg
\rightarrow c\bar{c}}$ that affects the values of $x_1$ and
$x_2$, the cross section and survival factor can be
calculated as before.  The results are shown in Fig. 15, which
exhibits a substantial degree of suppression at large $x_F$. 
Any data at large $x_F$ would put considerable constraint on
the models that attempt to explain the anomalous suppression
at small $x_F$.

\section{Conclusion}

In our attempt to understand the enhanced suppression of
$J/\psi$ at large $x_F$, we have found that the depletion of
gluons at large $x_1$ in the projectile is the most natural
explanation for the effect.  We have proposed an evolution
equation for the gluon distribution as the gluons propagate in
a nuclear medium.  The depletion at high $x_1$ contributes to
a mild growth of the gluon distribution at small $x_1$. 
However, that growth does not lead to any contradiction with
the existing data on $J/\psi$ suppression at mid-rapidity. 
Indeed, we have gone further to show where to find
informative clues on the dynamics of suppression at large
(positive and negative) $x_F$ in both $pA$ and $AB$
collisions.

What we have done here is only a modest first step towards
understanding parton evolution in nuclear matter.  While
concentrating on the gluons, we have ignored the influence of
the quark sector, a subject to be investigated at a later
point.  The depletion of quarks at large $x$ reveals itself in
the suppression of dilepton and leading meson production at
large $x_F$, the experimental evidences for which exist,
though in subtle ways.  Because of the conservation of Fermion
number, the degradation of the quark distribution at large $x$
cannot be substantial.  Nevertheless, the influence on the
gluon distribution at small $x$ may not be negligible.

An important implication of this work is that in $pA$ or $AB$
collisions the concept of a nucleon propagating through
nuclear matter as an identifiable, fixed entity needs
modification.  The usual notion that in nuclear
collisions the total transverse energy $E_T$ is proportional to
the number of nucleon-nucleon collisions would seem to have
difficulty in reconciling with the insistence that the nucleons
remain unaltered, if each inelastic collision of the nucleons
contributes a fraction of their energies to $E_T$.  The wounded
nucleon model \cite{bbc} makes a crude approximation of what goes
downstream as an average quantity that is different from the
incident nucleon, but ignores the way it changes as it
propagates.  Our evolution equation indicates that the parton
flux changes continuously and may emerge with a profile that
cannot be identified with that of a free nucleon in any sensible
comparison.  The revelation made by this understanding will
undoubtedly affect many aspects of high-energy nuclear
collisions.
 
\vspace{.5cm}  
\centerline {\large{\bf Acknowledgment}}
\vspace{.3cm}
We are grateful to Kari Eskola for providing us with the Fortran
codes for EKS98. This work was supported, in part, by the
U.S.-Slovakia Science  and Technology Program, the U. S. National
Science Foundation under Grant No. INT-9319091 and by the U. S.
Department of Energy under Grant No. DE-FG03-96ER40972.

\newpage
\begin{center}
\section*{Figure Captions}
\end{center}

\begin{description}
\item[Fig.\quad 1]\quad Nuclear shadowing and anti-shadowing
factor taken from EKS98 \cite{eks} for $Q^2 = 10$ GeV$^2$. 
The curves are fits by the simple formula in
Eq.~(\ref{9}).
\item[Fig.\quad 2]\quad The $\beta (\xi)$ function used to fit
EKS98 data points at $A = 100$.
\item[Fig.\quad 3]\quad $K_n$ calculated from the moments at
discrete $n$ and fitted by the formula in Eq.~(\ref{23}).
\item[Fig.\quad 4]\quad The $z$ dependences of $K_n(z)$.  The
lines are straight-line fits in the large $z$ region.
\item[Fig.\quad 5]\quad $Q_n$ as calculated from
Eq.~(\ref{25}).
\item[Fig.\quad 6]\quad The points of $Q_n$ are determined
from Eq.~(\ref{25}) or Fig.~5 at integer $n$ values; the
curve is a fit using Eq.~(\ref{29}). 
\item[Fig.\quad 7]\quad The kernal $Q(y)$ as calculated from
Eq.~(\ref{30}).
\item[Fig.\quad 8]\quad $g_n$ for two values of
$A$.
\item[Fig.\quad 9]\quad  The ratio of gluon distributions,
$G(x_1, A)$, for two values of $A$. 
\item[Fig. 10]\quad $\alpha (x_F)$ for $J/\psi$ production. 
The data points are from Ref.~\cite{ml}; the curves
represent our result.  
\item[Fig. 11] \quad $\alpha (x_F)$ for $\psi^{\prime}$
production.  The data points are from Ref.~\cite{ml}; the curves
represent our result.
\item[Fig. 12]\quad The survival probability for $J/\psi$
production in $pW$ collisions at different energies for the
entire range of $x_F$.
\item[Fig. 13]\quad The $z$ dependences of the coefficients
$a_i (z)$.
\item[Fig. 14]\quad The survival probability in $AB$
collisions.  The data points are from Ref.~\cite{mg}; the line
is composed of straightline sections connecting the
calculated points listed in Table 2.
\item[Fig. 15]\quad The survival probability for $J/\psi$ in
$Pb$-$Pb$ collision for all $x_F$ at two energies.

\end{description}
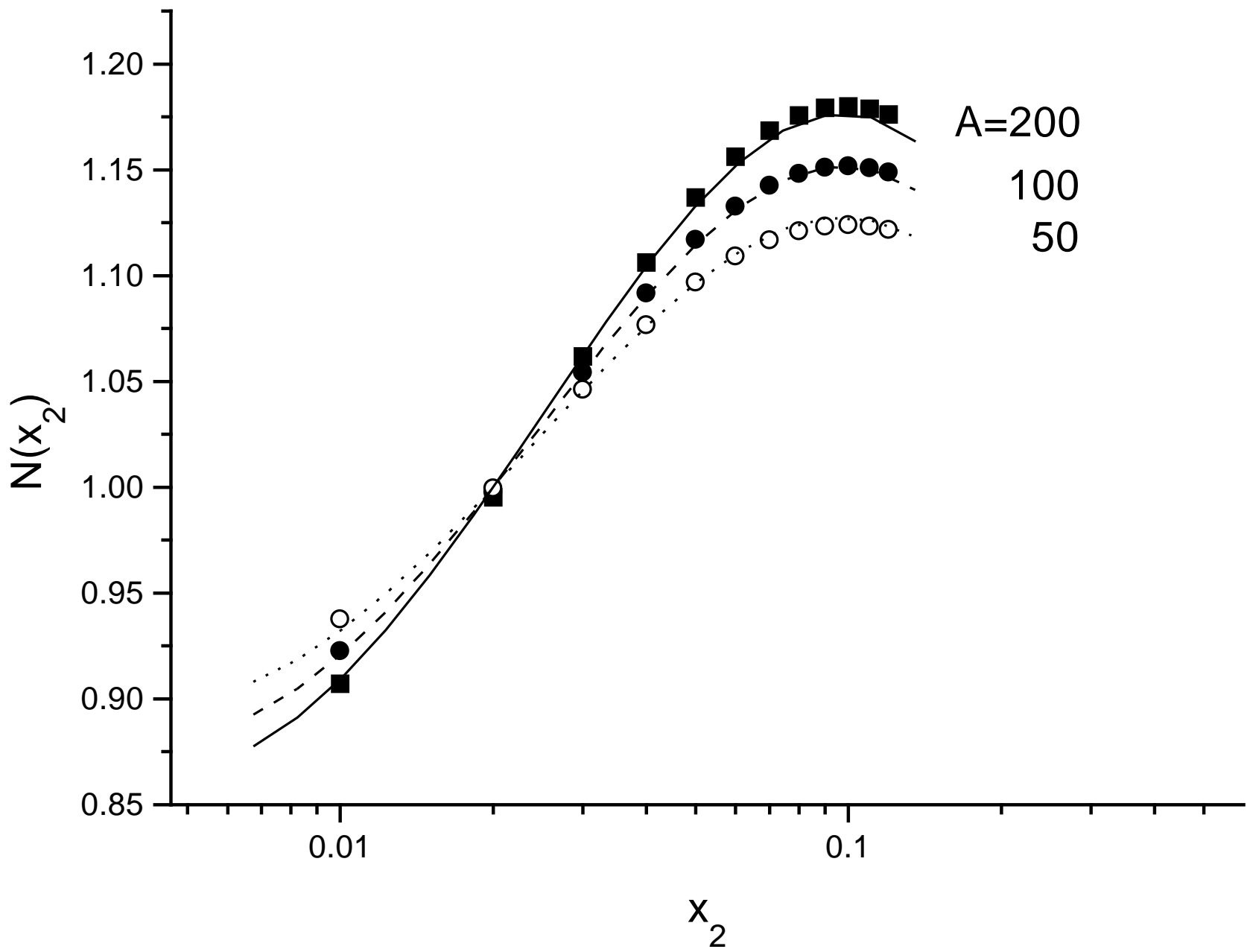
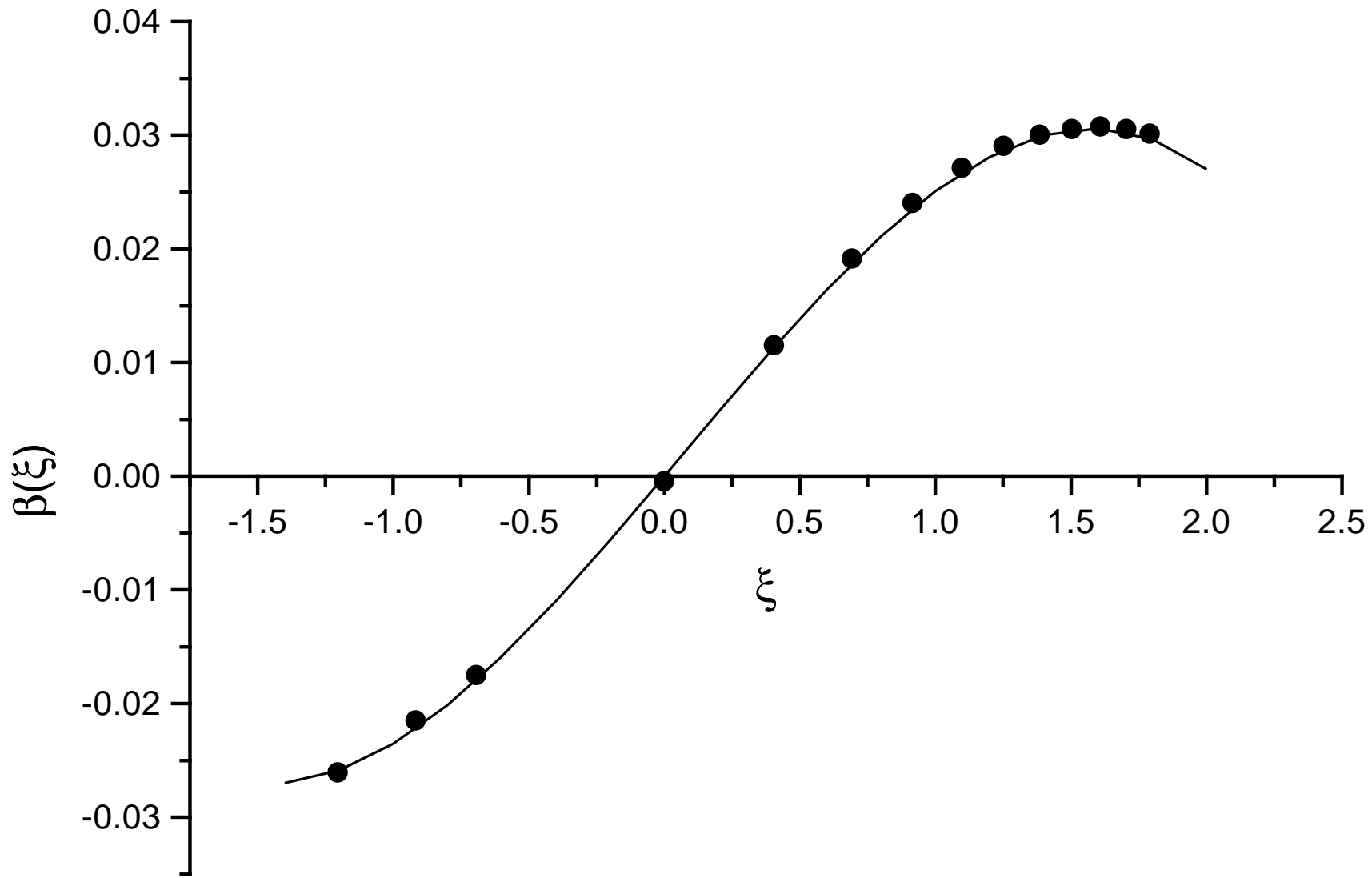
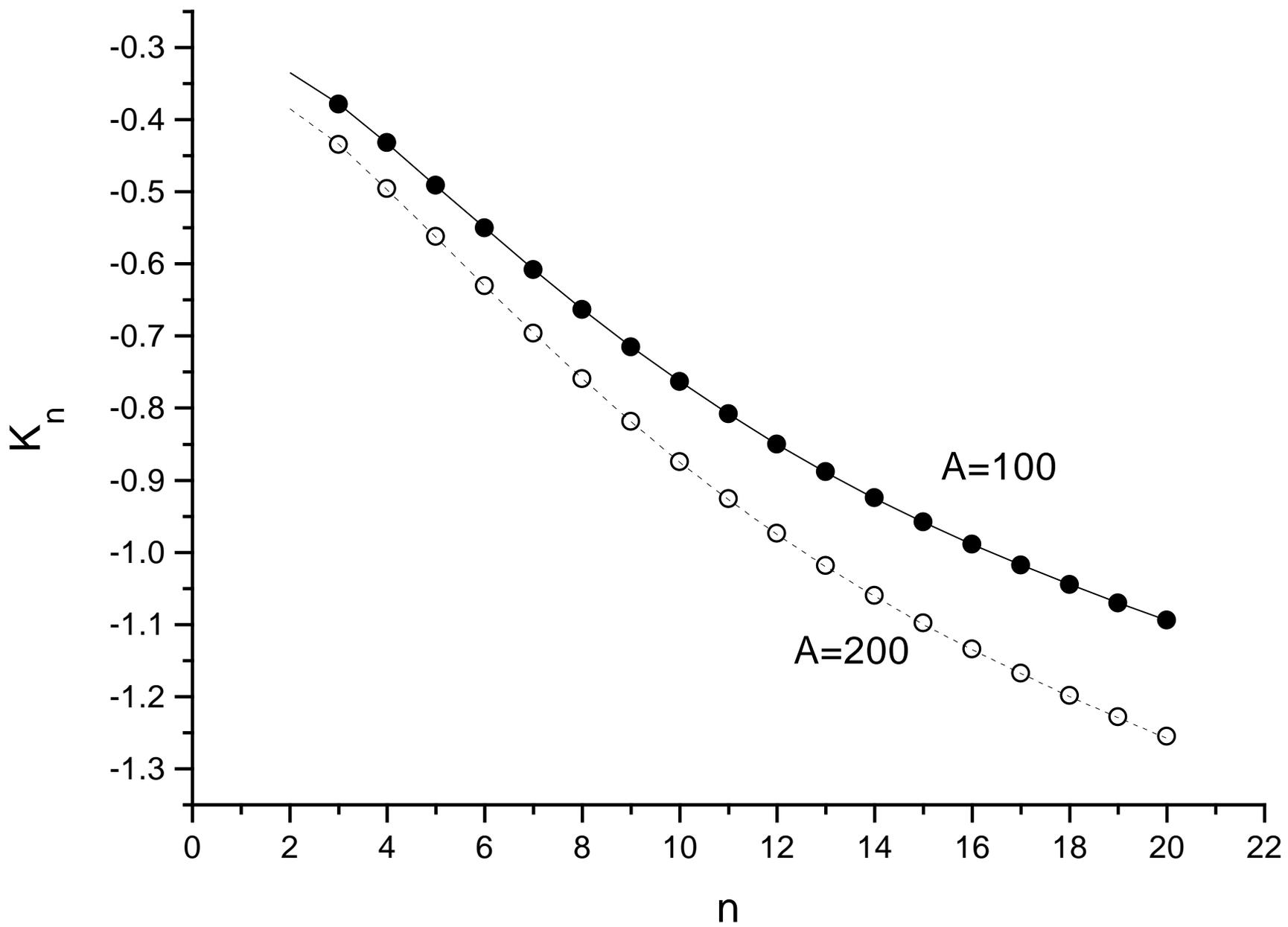
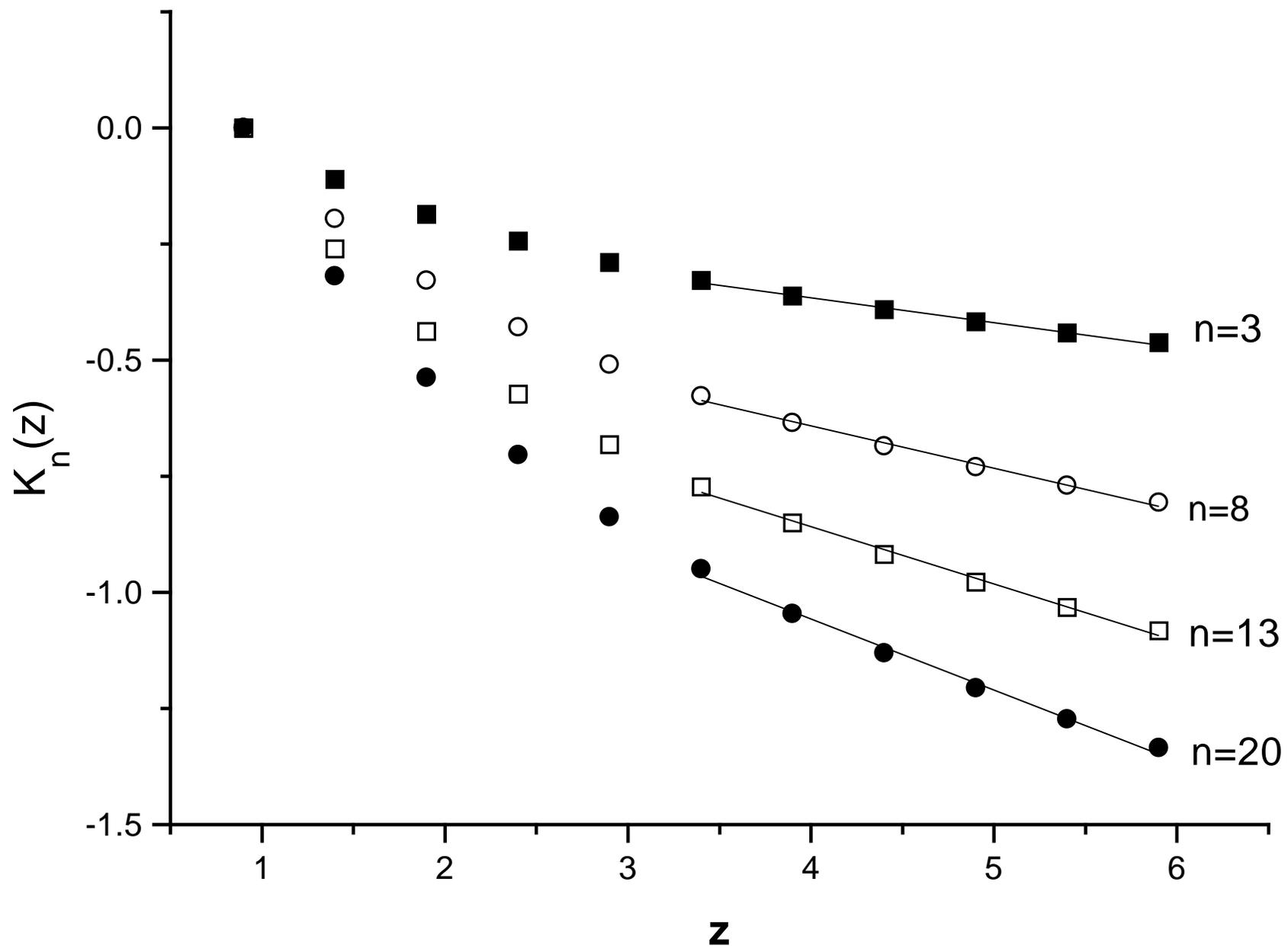
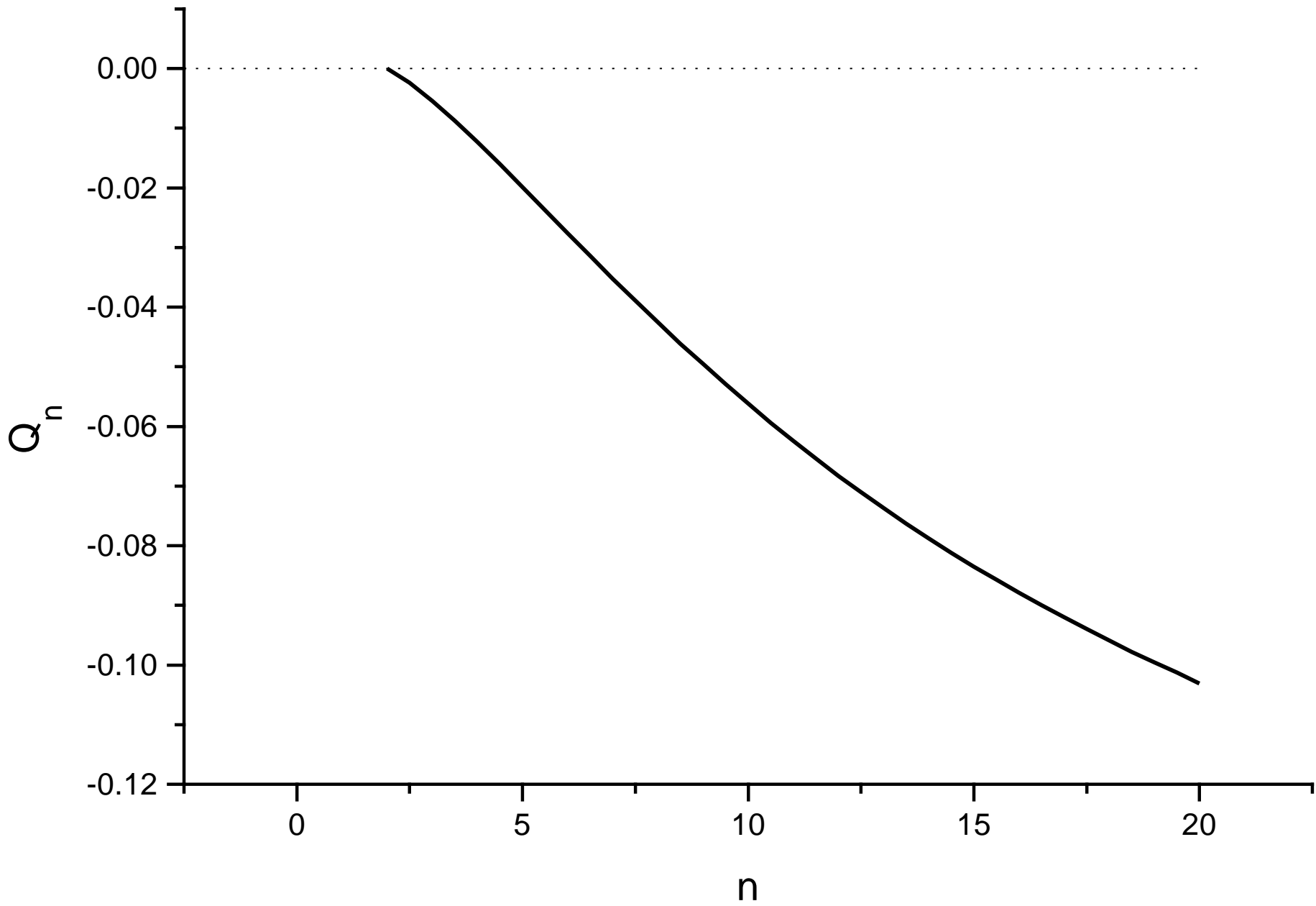
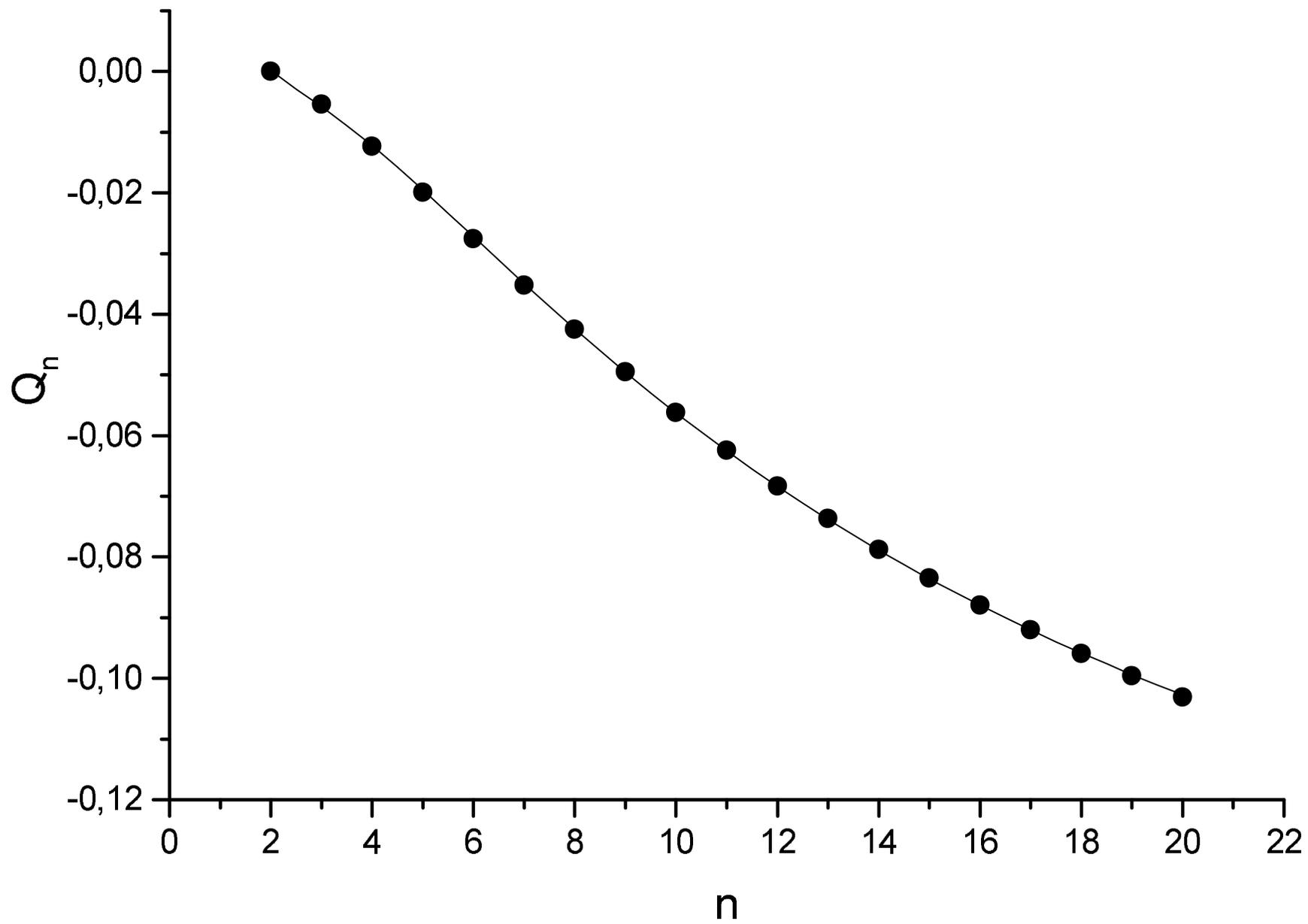
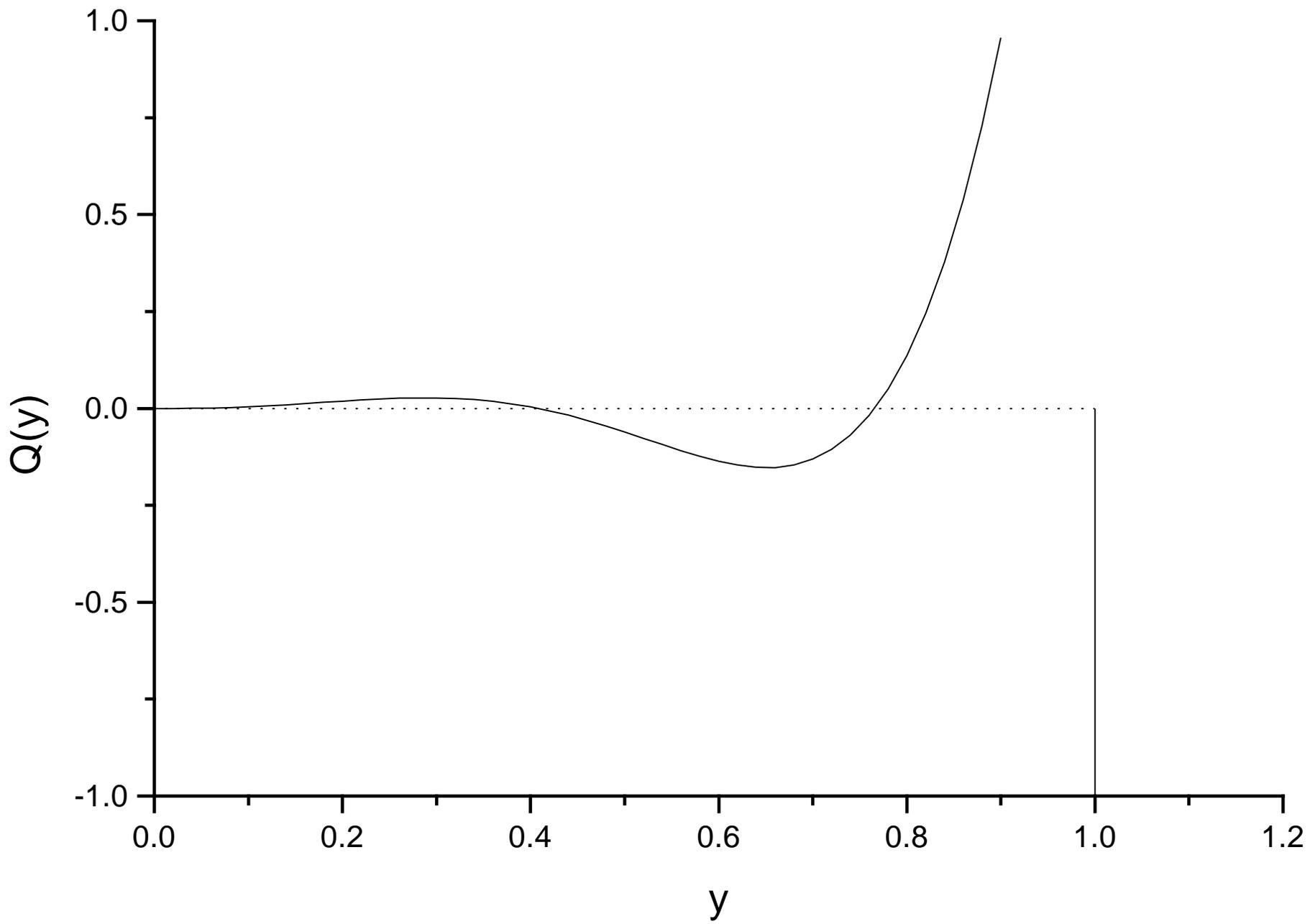
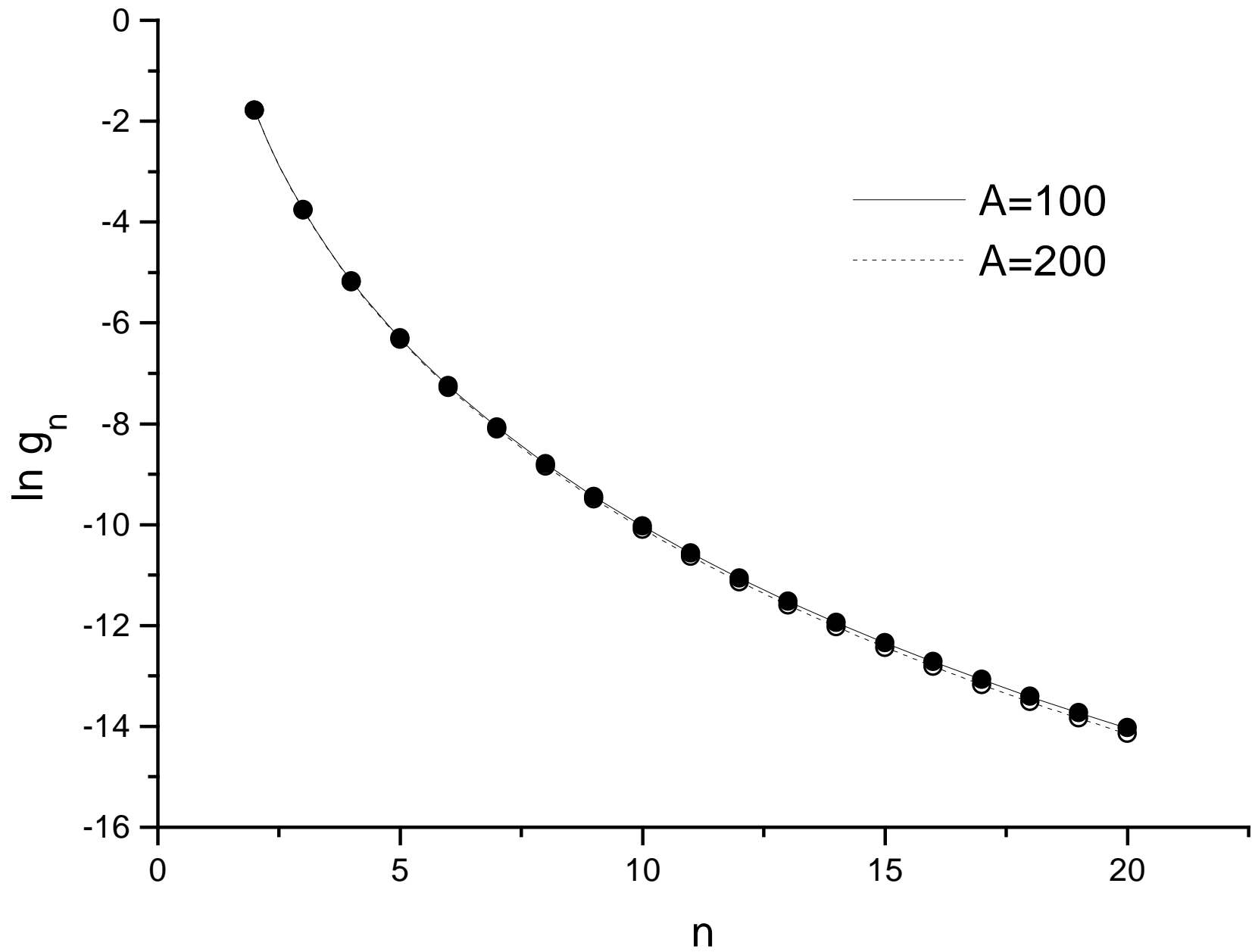
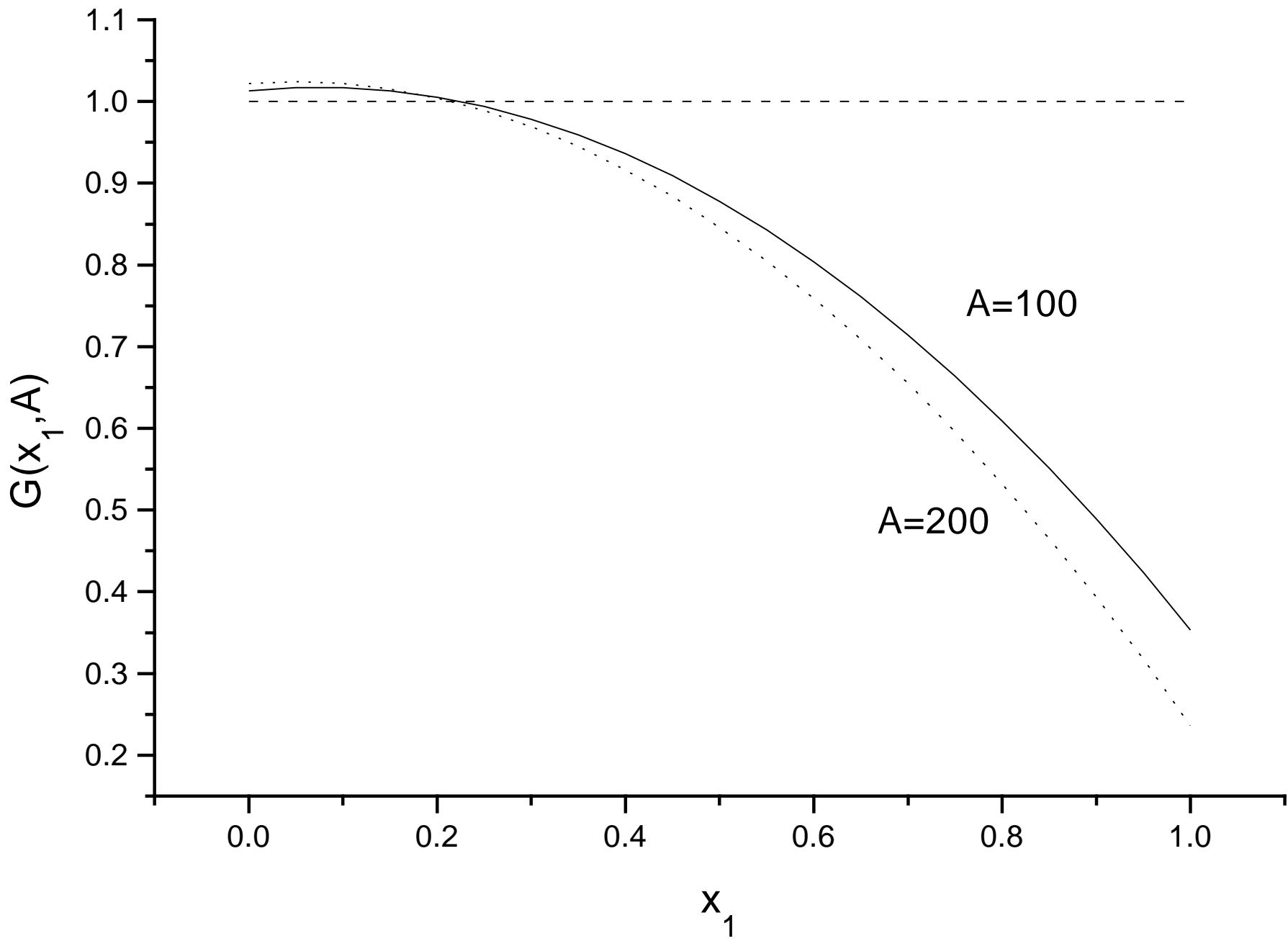
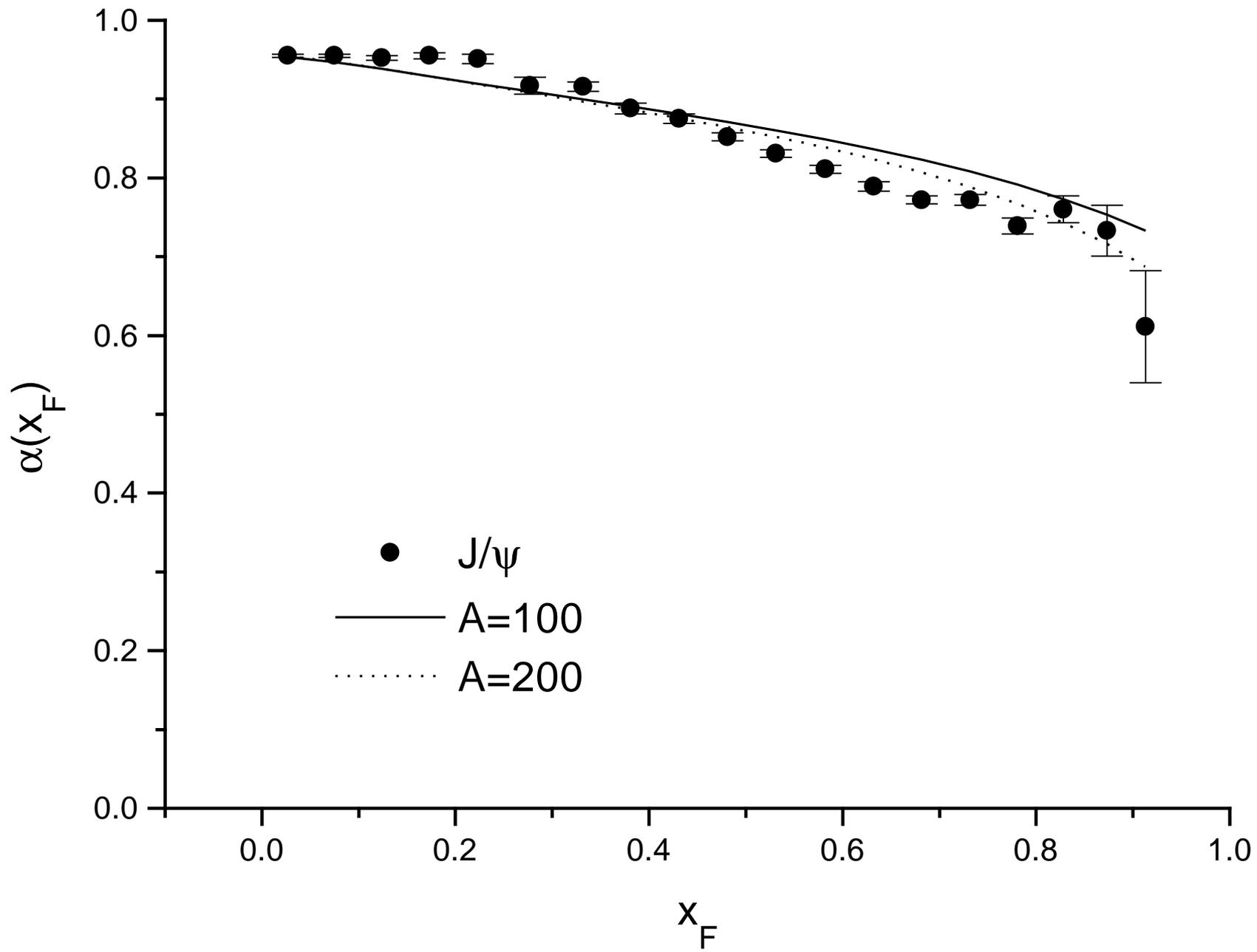
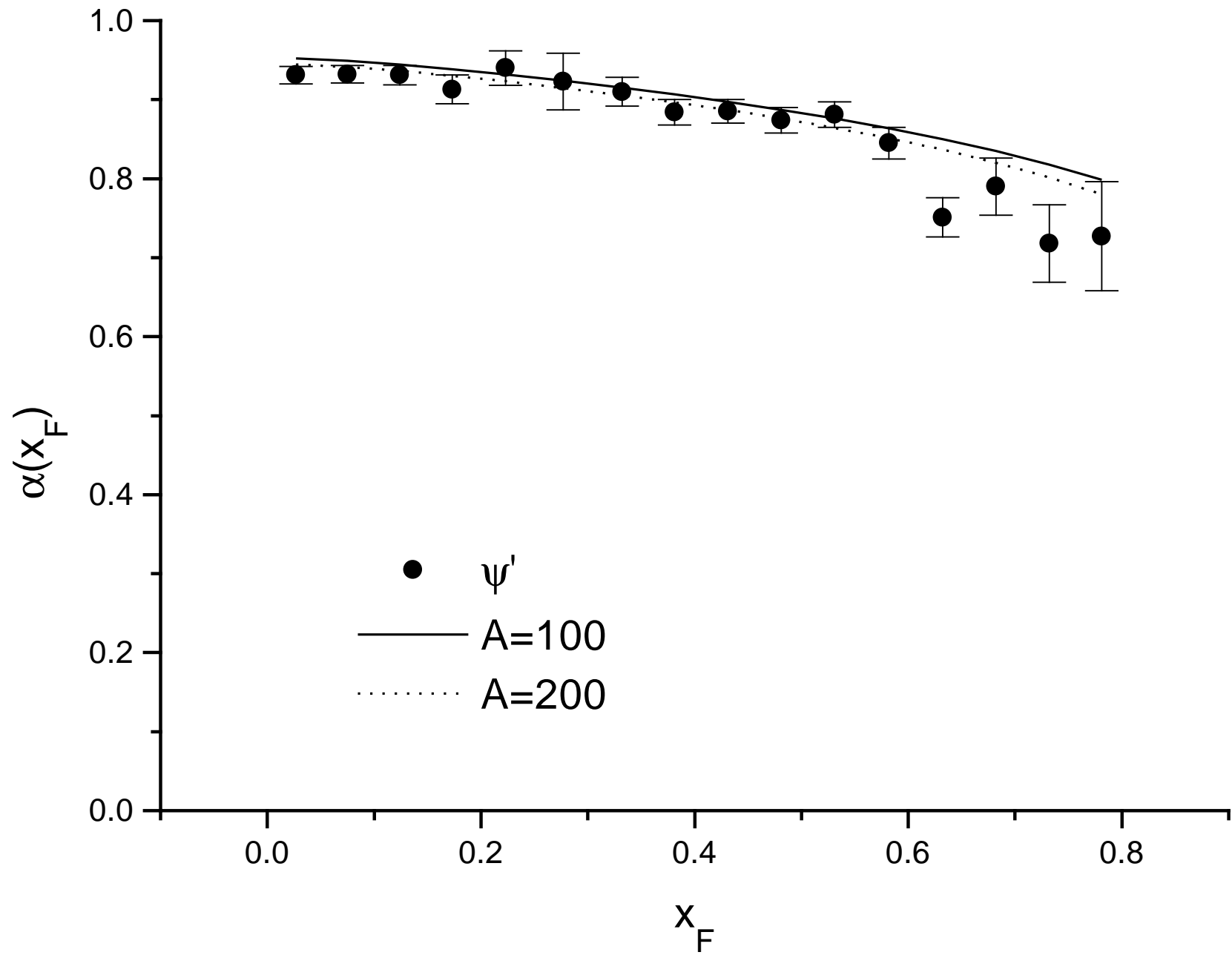
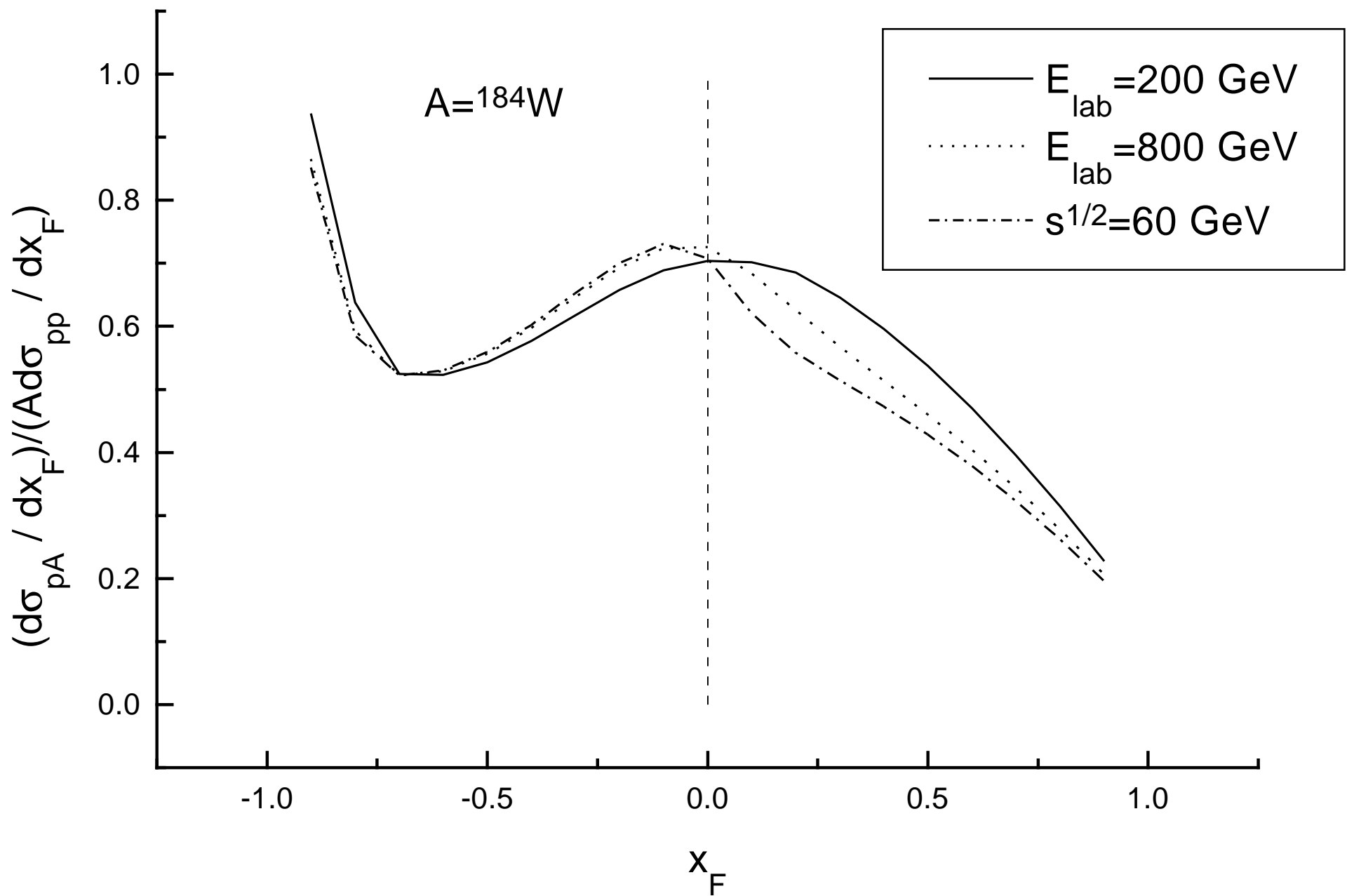
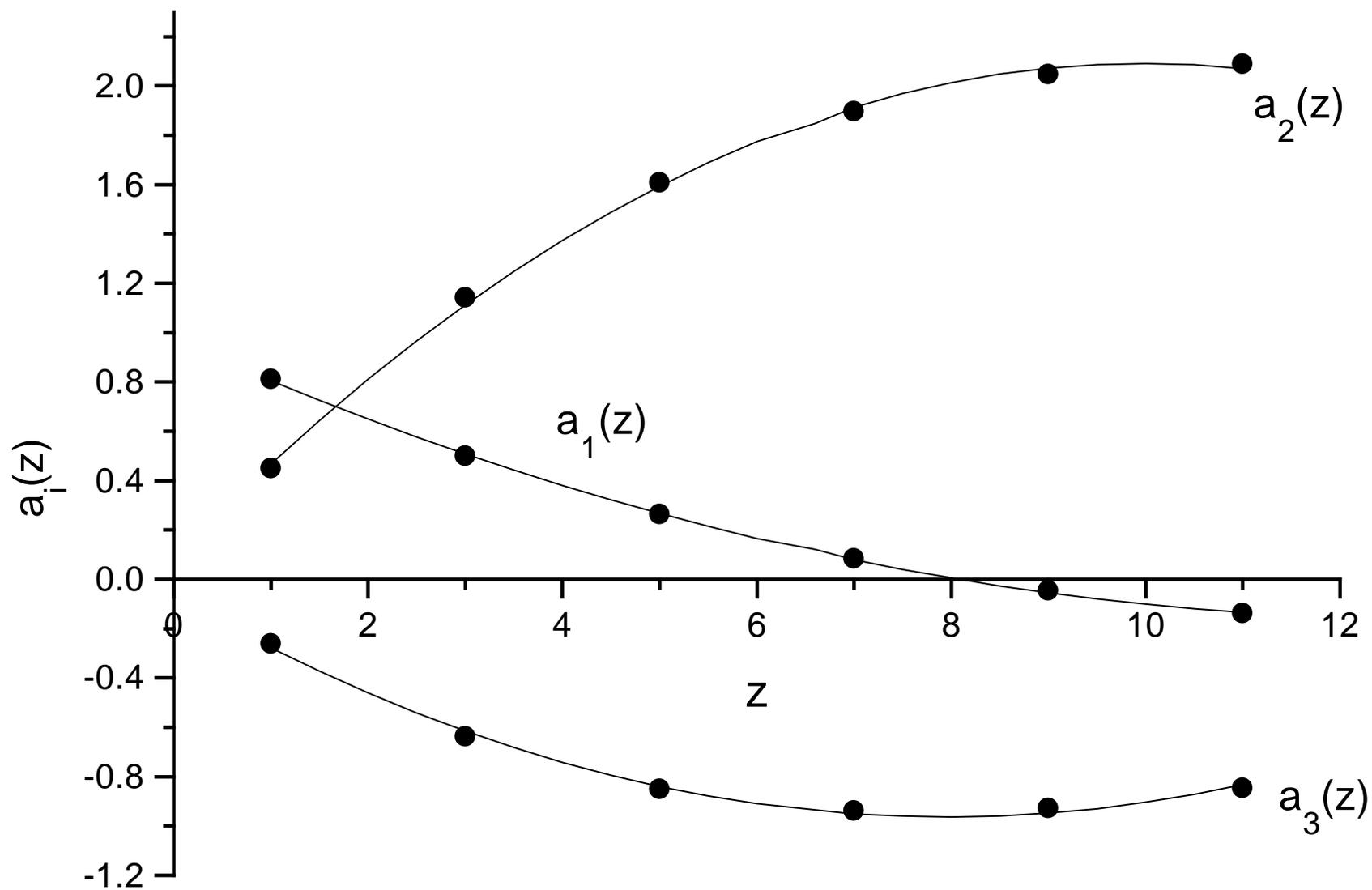
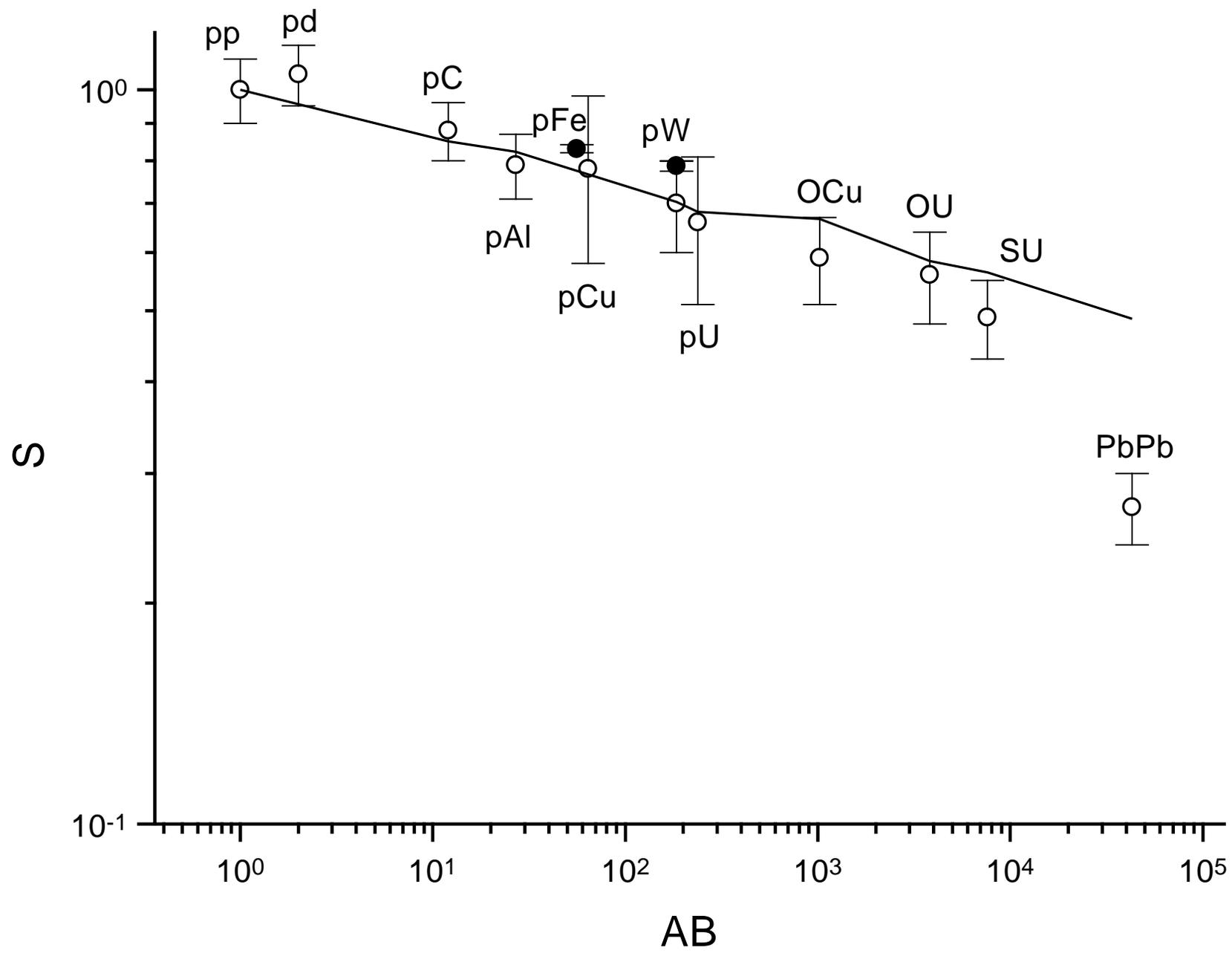
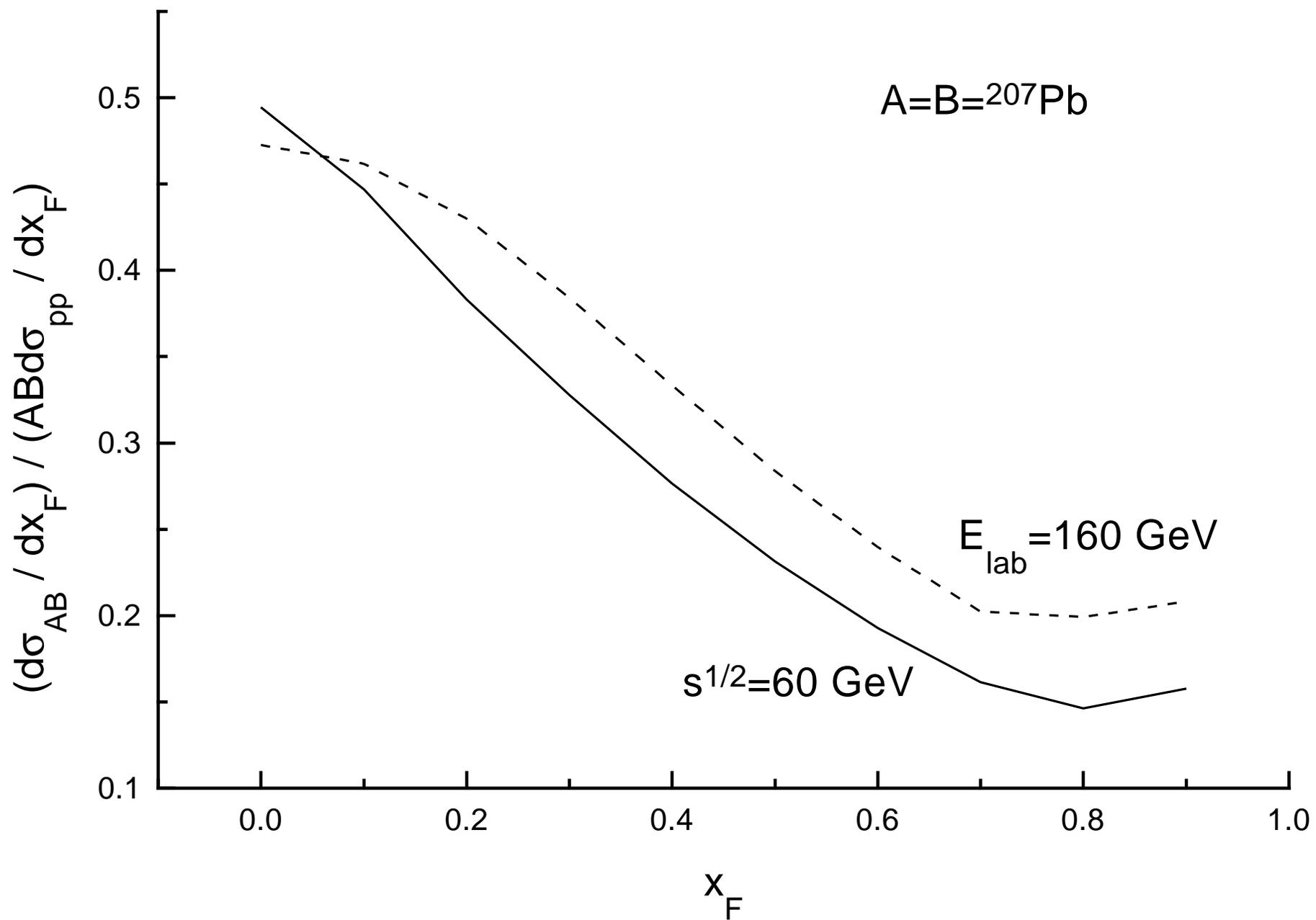
\end{document}